# LONG-RANGE RECONNAISSANCE IMAGER ON NEW HORIZONS

A. F. Cheng[1], H. A. Weaver[1], S. J. Conard[1], M. F. Morgan[1], O. Barnouin-Jha[1], J. D. Boldt[1], K. A. Cooper[1], E. H. Darlington[1], M. P. Grey[1], J. R. Hayes[1], K. E. Kosakowski[2], T. Magee[1], E. Rossano[1], D. Sampath[2], C. Schlemm[1], H. W. Taylor[1]

[1]*The Johns Hopkins University Applied Physics Laboratory, 11100 Johns Hopkins Road, Laurel MD 20723*
Tel. 240-228-5415
andrew.cheng@jhuapl.edu

[2]*SSG Precision Optronics, Wilmington MA*

## ABSTRACT

The LOng-Range Reconnaissance Imager (LORRI) is the high resolution imaging instrument for the *New Horizons* mission to Pluto, its giant satellite Charon, its small moons Nix and Hydra, and the Kuiper Belt, which is the vast region of icy bodies extending roughly from Neptune's orbit out to 50 astronomical units (AU). *New Horizons* launched on January 19, 2006 as the inaugural mission in NASA's New Frontiers program. LORRI is a narrow angle (field of view=0.29°), high resolution (4.95 μrad pixels), Ritchey-Chrétien telescope with a 20.8 cm diameter primary mirror, a focal length of 263 cm, and a three lens field-flattening assembly. A $1024 \times 1024$ pixel (optically active region), thinned, backside-illuminated charge-coupled device (CCD) detector is used in the focal plane unit and is operated in frame transfer mode. LORRI provides panchromatic imaging over a bandpass that extends approximately from 350 nm to 850 nm. LORRI operates in an extreme thermal environment, situated inside the warm spacecraft with a large, open aperture viewing cold space. LORRI has a silicon carbide optical system, designed to maintain focus over the operating temperature range without a focus adjustment mechanism. Moreover, the spacecraft is thruster-stabilized without reaction wheels, placing stringent limits on the available exposure time and the optical throughput needed to satisfy the measurement requirements.

# 1. Introduction

The *New Horizons* mission launched on January 19, 2006 on its way to perform the first reconnaissance of the Pluto-Charon system and the Kuiper Belt. First, however, New Horizons made a Jupiter swingby with closest approach on Feb 28, 2007. Extensive observations of the Jovian atmosphere, rings, and satellites were acquired. Pluto closest approach will occur on July 14, 2015. The prioritized measurement objectives of the New Horizons mission, and the contributions from LOng-Range Reconnaissance Imager (LORRI) observations, are summarized in Table 1.

Pluto is an icy dwarf planet with a significant atmosphere consisting mainly of nitrogen, CO and methane. High resolution images from LORRI will yield important information on Pluto's geology and surface morphology, collisional history, and atmosphere-surface interactions. Will Pluto have a young surface, with evidence of endogenic activity like plumes or cryovolcanism? Will there be evidence for tectonism in the form of faulting or ridge and groove formations? Will there be layered terrains? Will there be evidence for atmospheric hazes or for surface winds forming dunes (dunes on Pluto may be mostly grains of nitrogen ice)? Pluto is known to have an active surface, with changes in surface colors and reflectances observed by Earth-based telescopic monitoring. LORRI's high resolution images will reveal features as small as 100 m on Pluto (260 m on Charon).

Charon is Pluto's giant satellite: at about half the size of Pluto, it is larger than any other planetary satellite relative to its primary. Unlike Pluto, Charon has no detectable atmosphere, and it probably has an old surface that may preserve a cratering record from collisional evolution within the Kuiper Belt. LORRI data will play a critical role in determining the crater size distribution and morphologies on Charon. Equally important, LORRI images will provide precise measurements of the shapes and sizes of both bodies. In addition, LORRI will obtain high resolution images of the newly discovered moons Nix and Hydra (Weaver et al. 2005), comparable in terms of resolution elements across the illuminated disk to the Galileo images of Gaspra.

After the Pluto-Charon encounter, *New Horizons* will make the first visit to one or more Kuiper Belt objects. Owing to the likely small size of these targets, LORRI's high resolution is especially important to capture as much surface detail as possible. Will these Kuiper Belt objects look like the asteroid Eros, or will there be bizarre surface features like the flat-floored, steep-walled depressions (craters?) found on the nucleus of comet 81P/Wild 2?

The *New Horizons* mission has a long focal length, narrow angle imager for several reasons. Pluto is a dwarf planet, and *New Horizons* flies by quickly, so the encounter science observations occur within one Earth day – but with LORRI, *New Horizons* will be able to image the Pluto system at higher resolution than any Earth-based telescope can (even the Hubble Space Telescope, or its successor in 2015) for 90 days prior to encounter. These images will provide an extended time base of observations, for studies of the shapes, rotations, and mutual orbits of both Pluto and Charon, and for characterizing surface changes.

Moreover, Pluto and Charon both rotate at the same rate as for their 6.38 day mutual orbit, always keeping the same faces towards each other. Hence, during the near encounter, which lasts less than an Earth day, only one hemisphere of each body, that which faces *New Horizons*, can be studied at the highest resolution. The opposite faces of both Pluto and Charon are last seen some 3 days earlier, when the spacecraft is still ~4 million km away. Despite this distance, LORRI will obtain images with 40 km resolution. These will be the best images of the portions of Pluto and Charon which are not visible during the near encounter period.

Finally, we have not yet discovered the Kuiper Belt object(s) to which *New Horizons* will be targeted after the Pluto-Charon encounter, and extensive Earth-based observing campaigns are searching for potential targets. However, even after discovery, the heliocentric orbits of the targets cannot be measured with sufficient accuracy from Earth to enable the *New Horizons* spacecraft to fly to them, unless the targets are also observed directly from the spacecraft. The direction in which the target is seen from the spacecraft is then used to steer the spacecraft to the target by optical navigation. LORRI is expected to play a key role, by making the first and highest resolution detections of the Kuiper Belt target object from *New Horizons*, more than 40 days before closest approach.

The *New Horizons* instrument payload includes three imaging investigations: LORRI, the focus of this manuscript; Alice, a ultraviolet imaging spectrometer; and Ralph, a visible imager and infrared imaging spectrometer. These optical instruments are all approximately co-aligned to view a common boresight direction, and the spacecraft will maneuver as required to provide pointing during the various planetary encounters. The Ralph instrument includes a multispectral imaging channel (MVIC), which is a four-color, medium angle imager with a field-of-view (FOV) 5.7° wide, scanned by time-delay integration. LORRI provides complementary imaging data, as a panchromatic, narrow angle (field-of-view 0.29°) framing camera.

**Table 1. Pluto and Charon Prioritized Measurement Objectives**

| Group | Goal | LORRI Contribution |
|-------|------|--------------------|
| 1.1 | Characterize the global geology and morphology of Pluto and Charon | Hemispheric panchromatic maps of Pluto and Charon at best resolution exceeding 0.5 km/pixel |
| 1.2 | Map surface composition of Pluto and Charon | |
| 1.3 | Characterize the neutral atmosphere of Pluto and its escape rate. | Search for atmospheric haze at a vertical resolution <5 km |
| 2.1 | Characterize the time variability of Pluto's surface and atmosphere | Long time base of observations, extending over 10 to 12 Pluto rotations; panchromatic maps of the far-side hemisphere |
| 2.2 | Image Pluto and Charon in stereo | Panchromatic stereo images of Pluto and Charon |
| 2.3 | Map the terminators of Pluto and Charon with high resolution | High resolution panchromatic maps of the terminator region |
| 2.4 | Map the surface composition of selected areas of Pluto and Charon with high resolution | |
| 2.5 | Characterize Pluto's ionosphere and solar wind interaction | |
| 2.6 | Search for neutral species including H, $H_2$, HCN, and $C_xH_y$, and other hydrocarbons and nitriles in Pluto's upper atmosphere, and obtain isotopic discrimination where possible | |
| 2.7 | Search for an atmosphere around Charon | |
| 2.8 | Determine bolometric Bond albedos for Pluto and Charon | Panchromatic, wide phase angle coverage of Pluto and Charon |
| 2.9 | Map the surface temperatures of Pluto and Charon. | |
| 3.1 | Characterize the energetic particle environment of Pluto and Charon | |
| 3.2 | Refine bulk parameters (radii, masses, densities) and orbits of Pluto and Charon | Orbital parameters, bulk parameters of Pluto and Charon |
| 3.3 | Search for magnetic fields of Pluto and Charon | |
| 3.4 | Search for additional satellites and rings. | Search for satellites and rings; refine orbits, sizes, shapes of Nix and Hydra |

## 2. LORRI Requirements

LORRI is a long focal length imager (see **Figure 1**) designed for high resolution and responsivity at visible wavelengths. LORRI will perform its primary measurements while *New Horizons* approaches Pluto and its satellite Charon in July, 2015, obtaining images from a closest approach distance of 11100 kilometers to Pluto, at a fly-by speed of 13.77 kilometers per second. New Horizons performed a Jupiter flyby, with closest approach on Feb. 28, 2007. During the Jupiter system flyby, LORRI imaged the atmosphere of Jupiter, its ring system, and several of its satellites. Finally, after the Pluto and Charon encounter, the *New Horizons* spacecraft will be targeted to encounter a Kuiper Belt object (KBO), at which LORRI will again obtain high resolution images.

LORRI is required to obtain high resolution, monochromatic images under low light conditions. Other instruments on *New Horizons* have multispectral and hyperspectral capabilities. At Pluto encounter, 33 AU from the Sun, the illumination level is ~1/1000 that at Earth, but Pluto is an unusually bright object with a visible albedo of ~0.55. At the Kuiper Belt object, likely to be encountered outside 40 AU from the Sun, the illumination will be still lower, and the object will be darker, with an albedo of typically ~0.1.

The main objectives of LORRI are: (1) obtaining high resolution images of Pluto and Charon during the approach phase, including the hemisphere that will not be observed during closest approach, (2) taking images at closest approach with instantaneous field of view (IFOV) of approximately 50 meters per detector element and (3) obtaining optical navigation images required to support trajectory corrections. LORRI's reflective telescope is a Ritchey-Chrétien design, with a field-flattening lens group near the focal plane. The digital image is captured by a frame transfer CCD detector. The FOV is 0.29° x 0.29°; the 1024 x 1024 square detector elements have a pixel resolution (or IFOV) of 4.94 µrad. A summary of the characteristics of the LORRI instrument appears in Table 2.

**Table 2. Summary of LORRI Characteristics**

| |
|---|
| Visible Panchromatic Imager |
| Telescope Aperture 208 mm |
| Focal Length 2630 mm |
| Passband 0.35 – 0.85 µm |
| Field-of-view 0.29° × 0.29° |
| Instantaneous field-of-view 4.95 µrad |
| Back-thinned, frame transfer CCD |
| Nominal exposure times 50-200 ms |
| On-chip 4×4 pixel binning available |
| Autoexposure |

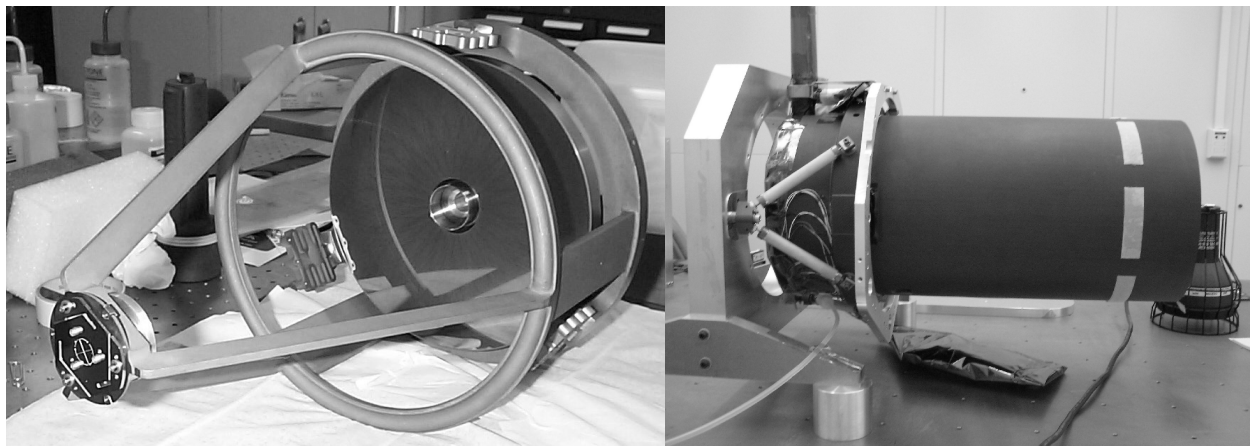

**Figure 1  (left) LORRI telescope assembly, showing SiC mirrors and metering structure; (right) LORRI composite baffle and flexure mount on test stand**

The LORRI instrument is mounted inside the *New Horizons* spacecraft, the interior of which is designed to remain near room temperature. As the telescope views cold space, and the CCD is designed to operate below −70° C, the thermal implementation of the system was a challenge. In order to maintain optical performance, a material with high thermal conductivity and low coefficient of thermal expansion (CTE) was required. As a result, the LORRI optical telescope assembly (OTA) has both primary and secondary mirrors and a metering structure fabricated from silicon-impregnated silicon carbide (see Figure 1).

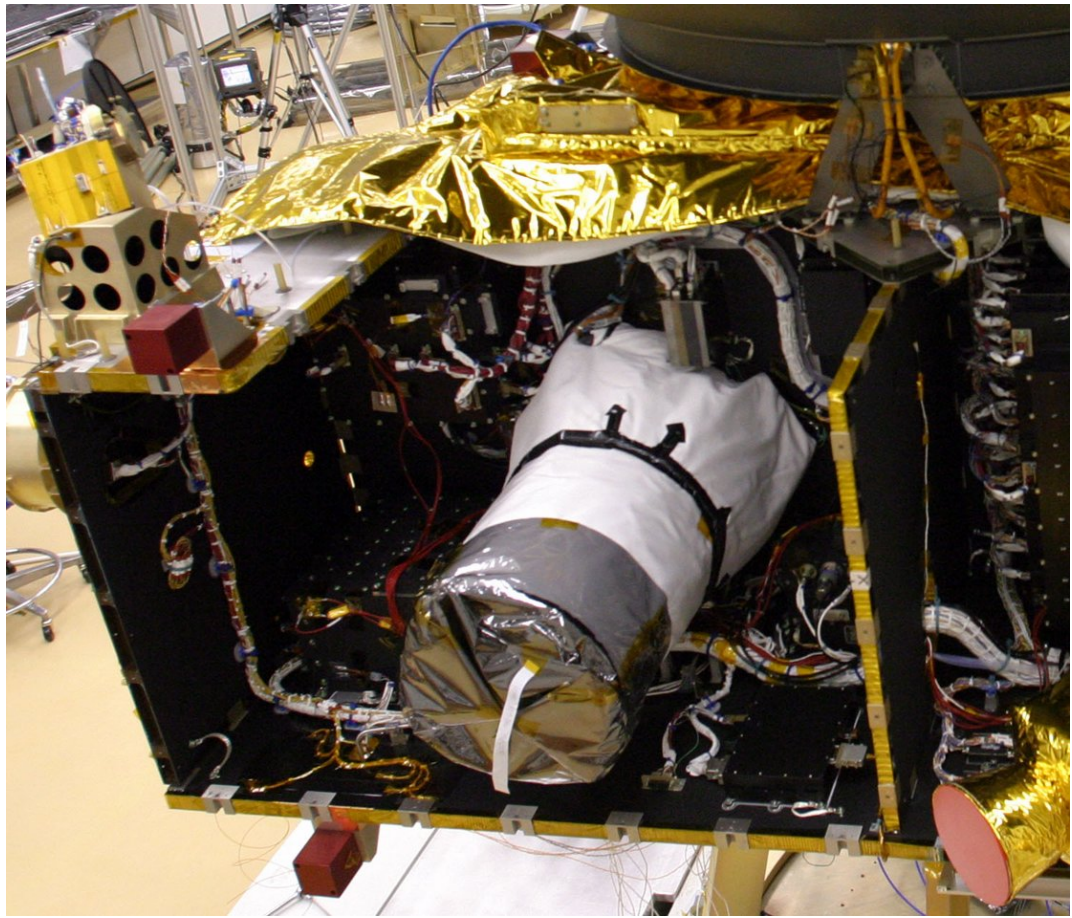

**Figure 2. LORRI mounted within New Horizons spacecraft, with outer panels of spacecraft removed. LORRI is wrapped in thermal blanketing, cover door closed.**

During the approach to Pluto, which occurs under high sun conditions (small phase angle or Sun-Pluto-spacecraft angle), LORRI is required to image the surface of Pluto at signal-to-noise ratio (SNR) > 100 in single frames. The encounter geometry is such that near closest approach to Pluto, where the highest resolution images would be obtained, LORRI views regions near the terminator under low sun conditions and still less illumination; here LORRI has the goal of imaging at SNR > 20 in single frames. Likewise, at the Kuiper Belt object LORRI has an SNR goal > 20.

The resolution requirement near Pluto closest approach is for LORRI to resolve 100 meters per line pair at a distance of 10,000 km from the surface. An IFOV required to be <5 μrad (Table 3) was derived, with a modulation transfer function (MTF) goal of 0.05 for spatial modulation at 1 cycle per 10 μrad. LORRI is designed to meet imaging requirements not only at nominal operating temperature (as low as approximately -100°C for the telescope), but also at room temperature.

For optical navigation, LORRI is required to be able to image a star of visual magnitude V=11.5 at SNR>7 in a single 100 ms exposure, with full width at half maximum (FWHM) >1 pixel. It is not desirable for too great a fraction of the energy from a point source to be imaged onto a single pixel, because stellar images become too undersampled. LORRI has a 4×4 pixel binning mode, for which its limiting magnitude requirement is V>17 in a single exposure of 9.9 s. This 4×4 pixel-binning mode will be used to search for the target KBO and to perform optical navigation on approach. A special spacecraft guidance mode is available for the KBO search in which the spacecraft will hold the target within the 4×4 pixel pointing tolerance for 10 second exposures. At 40 AU from the Sun, LORRI is predicted to be able to detect a 50 km diameter object, of albedo 0.04 and at phase angle 25°, from a distance of 0.35 AU, more than 40 days before the object would be encountered. This is ample time for targeting of the spacecraft.

**Table 3. LORRI Measurement Requirements at Pluto**

| Resolution | Resolve 100 m per line pair at 10,000 km |
|---|---|
| Derived requirement | Map full illuminated disk of Pluto at better than 1 km per pixel resolution with a 3×3 mosaic |
| Derived requirement | IFOV <5 μrad with FOV 0.29°; image frame pixel format 1024×1024 |
| Signal-to-Noise Ratio | Achieve SNR >100, single pixel, single exposure at Pluto during approach, for albedo 0.55 at low phase angle |
| Derived requirement | Achieve SNR >20 (goal), single pixel, single exposure at Pluto, at 110° phase angle near terminator |
| Optical Navigation | Achieve SNR >7 for star of visual magnitude 11.5 with a single exposure of 100 ms |
| Derived requirement | Achieve stellar limiting magnitude >17 for 9.9 second exposure in 4x4 binned mode for KBO search at SNR >7 |

Observations of the Pluto system begin at least 90 days prior to the July, 2015 encounter, with both Pluto and Charon already resolved (see Table 4). Initial observations are planned to refine the orbits of Pluto, Charon and the two newly discovered moons Nix and Hydra. Pluto and Charon can be imaged in single frames or in 2×1 mosaics through at least ten full orbits (orbit period 6.38 days) ending about 14 days before Pluto closest approach (c/a), to refine orbital ephemeredes and especially the eccentricity. In the last week prior to c/a, searches for librations of Pluto and Charon are performed, where Pluto subtends 123 pixels one half rotation before c/a. Nix and Hydra are expected to be detectable in 4×4 binned images about 90 days before c/a, and in unbinned single frames within the last 14 days.

The last full frame image of Pluto is acquired about 10 hours before c/a, and two 3x3 global mosaics of the full illuminated disk are acquired during the near encounter. The full illuminated disk observations will be useful to construct global base maps, to determine the global shape of Pluto, and to search for oblateness and tidal bulges. Additional images near c/a are obtained at successively higher resolutions but covering smaller portions of the illuminated disk. The Pluto near encounter image dataset will be used to study surface morphology, geologic processes, and atmosphere-surface interactions. Likewise, the full illuminated disk of Charon will be imaged with 3x3 LORRI mosaics at resolution of better than 0.5 km per pixel, meeting the Group I panchromatic imaging requirement for Charon (the corresponding 0.5 km per pixel imaging requirement for Pluto will be met by MVIC).

**Table 4. LORRI Planned Observations of the Pluto System**

| |
|---|
| Approach imaging starts c/a-90 days, Pluto and Charon already resolved, Nix and Hydra detectable (refine orbits) |
| Through at least ten full orbits ending c/a-14 days, both Pluto and Charon can be acquired in single frame or 2x1 mosaic together with background stars |
| At c/a-14 days, Pluto subtends 28 pixels |
| By c/a-7 days Pluto subtends 57 pixels, both Pluto and Charon can be acquired with 3x1 image strip |
| At c/a-3.2 days, Pluto subtends 123 pixels, imaging far side of Pluto (unseen side during c/a) |
| At c/a-10 hours, last full frame image of Pluto, 2.5 km/pixel at nadir |
| Near encounter Pluto observations – 3x3 mosaics of full disc better than 1 km/px; image strips at progressively finer resolutions to 110 m/px; terminator imaging sequence with 50 m/px resolution (goal) |
| Near encounter Charon observations – full frame illuminated disk images; 3x3 mosaics of full disc better than 0.5 km/px; terminator observation 130 m/px |
| Near encounter observations of Nix and Hydra, better than 200 m/px resolution (goal) |

The LORRI data sets acquired during the Jupiter encounter in January through March of 2007 are summarized in Table 5. Images of the Jovian atmosphere and its clouds and storms include full disk rotation sequences, acquired up to two months before Jupiter c/a on February 28, as well as high resolution 2×2 mosaics of specific features acquired near c/a. Imaging sequences were executed for each of the Galilean satellites, including observations of the night sides of Io, Europa, and Ganymede acquired while the respective satellites were in eclipse (i.e., in Jupiter's shadow). The Jupiter ring system was imaged at both low and high solar phase angles, as well as during the ring plane crossing. Images to study the shapes and photometric properties of Elara and Himalia, two of Jupiter's irregular satellites, were obtained.

**Table 5. LORRI Observations of the Jovian System**

| |
|---|
| Global imaging of Jupiter atmosphere, full rotation sequences, during approach |
| Near encounter imaging of Jovian cloud and storm dynamics near c/a, best resolution 12 km/pixel |
| Global imaging of Io, illuminated portion, best resolution 12 km/pixel |
| Io plume inventory (plumes higher than 60 km) |
| Io nightside imaging from eclipse; Io hot spots and auroral emissions |
| Global imaging of Europa; map broad, regional-scale arctuate troughs, best resolution 15 km/pixel |
| Map Europa nightside auroral emissions in eclipse |
| Map Ganymede nightside auroral emissions in eclipse |
| Global imaging of Callisto, best resolution 23 km/pixel |
| Jovian ring plane crossing; map vertical structure of ring systems |
| Map longitudinal structure of Jovian rings |
| Resolved images and phase curves of irregular satellites Himalia and Elara |

# 3. LORRI Instrument Description

## 3.1   LORRI Overview

The LORRI instrument was designed and fabricated by a combined effort of The Johns Hopkins University Applied Physics Laboratory (JHU/APL) and SSG Precision Optronics Incorporated (SSG), of Wilmington, Massachusetts, USA. LORRI has four subassemblies in close proximity connected by electrical harnesses. These are the OTA, the aperture cover door, the associated support electronics (ASE), and the focal plane unit (FPU). Except for the door, all are mounted inside the spacecraft on its central deck; the door is mounted to an external spacecraft panel. LORRI is electronically shuttered, with no moving parts aside from the cover door. The ASE implements all electrical interfaces between LORRI and the spacecraft except for the door control, several spacecraft thermistors, and two decontamination heaters. Figure 3 is a block diagram of the instrument. Conard et al. (2005) give a detailed description of LORRI design, manufacture and test.

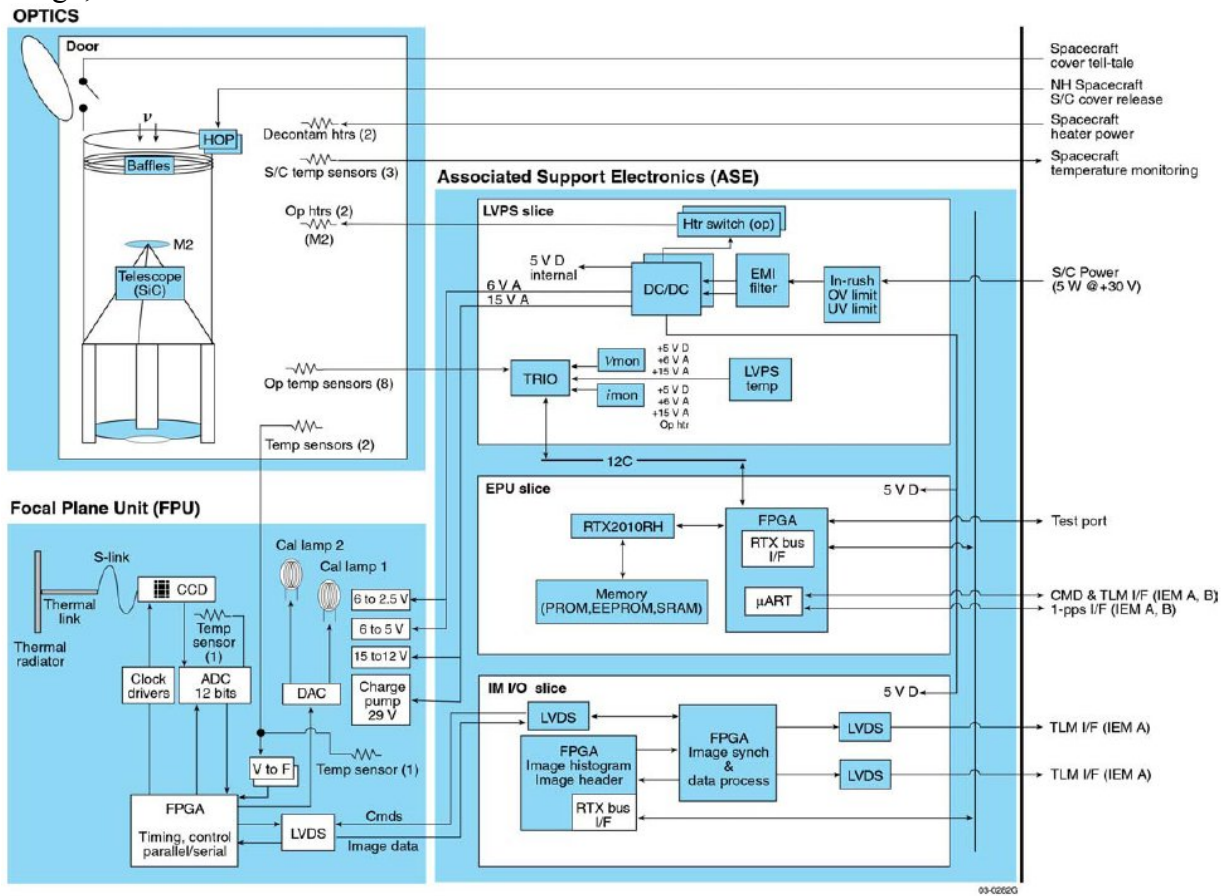

**Figure 3. LORRI block diagram , showing subassemblies: optics (OTA) and aperture door; focal plane unit (FPU); and associated support electronics (ASE) with three slices, which are the low voltage power supply, the event processing unit, and the imager input/output board.**

A summary of LORRI instrument specifications is given in Table 6. The LORRI boresight is required to be aligned within 0.1° of the boresight of the Ralph imager (MVIC). For a combined summary of the fields-of-view for the three imaging instruments on New Horizons, see the payload overview companion paper (Weaver et al. 2007).

The OTA, aside from the focal plane unit and thermal blanketing, was designed and built by SSG Precision Optronics, Inc. The primary and secondary reflecting optical elements are constructed of SiC. The telescope is a 2630 mm focal length, f/12.6 Ritchey—Chretien design. Three field-flattening fused silica lenses, located in front of the focal plane unit, are the only refractive elements in the system.

**Table 6. LORRI Instrument Specifications**

| |
|---|
| Optical telescope assembly mass 5.6 kg |
| Total mass 8.6 kg |
| Electrical Power 5 W |
| Heater Power 10 W |
| Focal plane calibration lamps (two) |
| Data interfaces Low Voltage Differential Signaling (LVDS) and RS-422 (both dual redundant) |
| ADC 12 bit |
| Image format 0: 1024×1028 (including 4 dark columns) |
| Image format 1: 256×257 (4x4 binned, including one dark column) |
| Embedded image header (first 408 bits of image data per frame, either format) |
| 32-bin image histogram provided for every image |

The SiC metering structure of the telescope holds the mirrors and field flattening lens cell. It is a monolithic structure consisting of a primary mirror (M1) bulkhead, short cylindrical section, and three-blade spider with secondary mirror (M2) mounting. The field flattener assembly mounts to the M1 mounting plate and protrudes through the M1 mirror. The metering structure is mounted to the graphite composite baffle using three titanium, vibration-isolating feet. The baffle assembly is mounted to the spacecraft using six glass-epoxy legs, which provide thermal isolation. The entire OTA is covered with multi-layer insulation (vented away from the OTA), except for the entrance aperture.

LORRI is protected from contamination and solar illumination using a one-time-open door mechanism. The door is mounted to the exterior of the spacecraft, and the LORRI baffle tube extends into the door to form a contamination seal. The door is aluminum, with thermal blankets for temperature control prior to deployment. The mechanism uses redundant springs and redundant paraffin actuators for deployment. A port allowing for installation of a witness mirror or small window is also part of the door. The door was opened successfully in flight on August 29, 2006.

The ASE provides the data and control interfaces to the spacecraft and to the focal plane unit. It consists of three 10 cm by 10 cm printed circuit cards electrically interfaced to one another via stackable connectors. They are in a magnesium housing, mounted directly to the spacecraft deck, a short distance from the OTA.

The LORRI FPU has a back-illuminated, thinned, high quantum efficiency CCD (an E2V Technologies Model 47-20). The FPU consists of a magnesium box mounted to the spacecraft deck, housing a 15 cm by 10 cm circuit card that controls the frame transfer CCD and provides interfaces to the imager board of the ASE. This circuit card is connected by a flex circuit to a small electronics board mounted at the focal plane. The small focal plane board holds the CCD itself and is mounted on thermal stand-offs, allowing the CCD and the small board to operate at ≤ -70° C while the magnesium box operates near room temperature. The CCD is in a window-less mount to avoid scatter and multiple reflections, with a black, anodized aluminum plate installed over the CCD storage readout area.

### 3.1.1 Design requirements and trades

The stringent optical, thermal and structural requirements for the LORRI OTA presented many design challenges. The primary design driver for LORRI instrument was the resolution requirement as described in section 2.2. The resolution is limited by a number of factors, including the stability of the spacecraft while an image is being exposed. The pointing stability of the spacecraft is characterized as a typical drift rate of 25 μrad per second. The minimum exposure time is limited by the frame transfer time of the CCD. In order to remove completely the image smear which occurs during transfer, the exposure time should preferably exceed the frame transfer time (approximately 13 milliseconds), although acceptable image quality has been achieved in flight at exposures as short as 1 millisecond. LORRI was designed for an exposure time range between 50 and 200 milliseconds, with 100 milliseconds the nominal design value. Over this exposure range, the spacecraft orientation would drift ~2 to ~7 μrad.

After the range of exposure times was determined, IFOV had to be traded. A smaller IFOV yielded higher resolution, though it would ultimately be limited by spacecraft stability. Diffraction limited system resolution when the entrance pupil became appreciably smaller than 200 mm diameter. Strict mass limitations combined with cost limitations prevented increasing the aperture much beyond 200 mm diameter. An aperture of 208 mm was selected.

With reasonable assumptions about the type of telescope and detector and their associated efficiencies, and with the range of nominal exposure times, a nominal IFOV of 5 μrad was chosen, which defined an effective focal length of 2630 mm. The FOV of the final design was 0.29° square.

The aperture requirement drove the telescope to reflecting optics. Mass and cost limitations, combined with the FOV and imaging requirements, drove the design to a Ritchey-Chrétien design. Refractive elements were used as field flattening lenses, as the Ritchey-Chrétien focal plane curvature over the flat CCD would have limited imaging performance without them. The field flattening lenses allowed LORRI to meet the requirement of <0.1% distortion at all points over the full field of view. There was no requirement for color imaging. An optical ray trace layout for LORRI is shown in Figure 4.

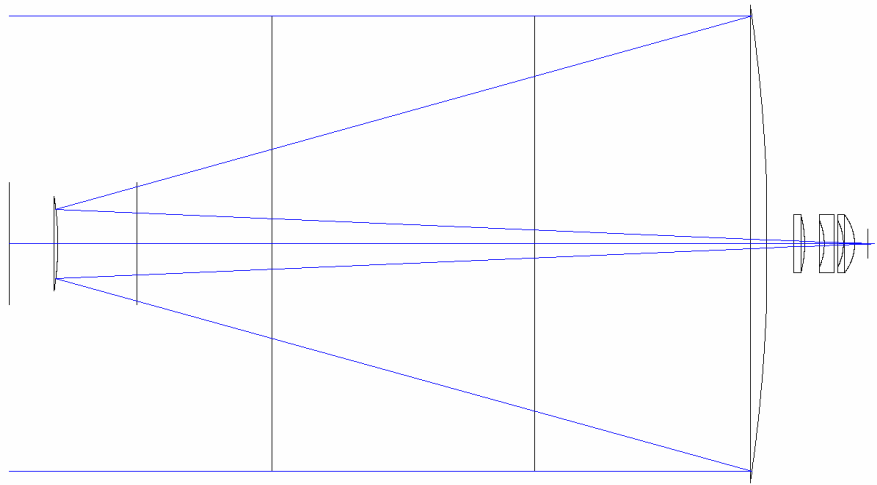

**Figure 4. Optical Ray Trace Layout, showing primary and secondary mirror surfaces, plus fused silica lenses.**

During the New Horizons mission, LORRI is exposed to the following radiation environments: a total dose of ionizing radiation 15 krad and a neutron fluence (worst case) of 1.4 × 10$^{10}$ n/cm$^2$. The neutron fluence is from the radioisotope thermal generator over the course of the mission. In the OTA, the refractive optics, antireflective coatings, and reflective coatings were required to be tolerant to the specified radiation environments. The LORRI CCD is operated below -70°C to mitigate degradation of charge transfer efficiency from exposure to the neutron fluence.

### 3.1.2 Optical design

The LORRI OTA is a Ritchey-Chrétien design, with high system throughput required because of the short allowed exposure time and low light level at Pluto. The complete LORRI OTA design was evaluated with a computer-aided design model including stray light analysis. Specular reflections and bidirectional reflectance distribution functions of the Aeroglaze Z-306 black paint, primary and secondary mirrors, field group optics and focal plane were included in the model. The primary and secondary baffle tubes were sized to minimize obscuration and suppress direct paths to the FPA's active area. The telescope magnification and obscuration were balanced, affecting the optical sensitivity and MTF, respectively. Out-of-field stray light was evaluated by generating point source transmittance curves with angular scans across the boresight in two orthogonal directions (-70° to +70° for each scan) to search for any obscured paths with unacceptable amplitude.

The OTA design required multiple baffle vanes, fabricated from graphite composite material, surrounding the metering structure to suppress stray light. In addition, an inner baffle is used extending from the hole in the primary mirror. This inner baffle has both inner and outer vanes plus threading in its interior. All baffle design features were optimized through TracePro ray tracing analysis. This analysis shows that out-of-field stray light is adequately suppressed and that ghosting is acceptable. The derived OTA system root-mean-squared (rms) wavefront error requirement based on the MTF requirement was <0.10 waves @ 632.8 nm, over an operating environmental temperature range of -125°C to 40°C and after g-release. The system throughput requirement drove a surface roughness requirement on all optics to a goal of <20 Å rms.

### 3.1.3 Thermal requirements and design

The extreme thermal environment posed a design challenge. LORRI is mounted within the interior of a spacecraft with deck temperature in the range 0 to +40ºC temperature, while it has an open, 208 mm aperture radiating into cold space. The challenge was to minimize defocus due to gradients developed in the OTA metering structure. The derived requirement was to limit the heat loss from the spacecraft to the OTA to less than 12 W, requiring conductive isolation at the interface and near zero radiative coupling to the spacecraft. The gradients in the OTA metering structure were limited to 2.5° C axial and 1.0° C lateral, and 0.5 W heater power was available for mirror gradient control without any in-flight focus adjustment mechanism.

The thermal requirements were met by an athermal, single material solution, which is self-compensating under soak conditions. SSG's SiC 55A formulation was chosen due to its inherent high conductivity, which acts to minimize gradients, and its low coefficient of thermal expansion, which minimizes the thermal strain impact of such gradients. Invar 36, a good match to SiC 55A over the temperature range of interest, was chosen for the metallic inserts that allow bolting together of the OTA assembly. All invar inserts, as well as the secondary mirror foot, were epoxy bonded to the SiC. In addition, the OTA is mounted inside of the telescope baffle tube, made of highly conductive K13C graphite composite. The baffle tube provides a relatively uniform cold sink along the length of the telescope which helps to reduce longitudinal thermal gradients.

Additional thermal control features on the LORRI telescope help to reduce system gradients. The telescope is mounted to the spacecraft via G-10 isolators which isolate the OTA conductively from the spacecraft deck. Covering the appropriate OTA and spacecraft surfaces with multi-layer insulation minimizes radiative coupling between the OTA and the spacecraft deck. The LORRI multi-layer insulation, which represents 15% of the instrument mass, consists of 23 separate pieces.

Areas where material mismatches occur are flexured or otherwise configured to minimize induced strain and resulting optical degradation. The secondary mirror is mounted to a flexured Invar 36 mount plate, which in turn mounts to the end of the SiC metering structure. The secondary and primary mirror magnesium baffles are flexure mounted to the OTA structure (SiC) and the lens cell (Invar 36), respectively. The system aperture stop is made of aluminum 6061-T6 and is flexure-mounted to the middle ring of the OTA structure. The CCD mount plate is attached to the OTA structure by titanium flexures which have the dual purpose of mitigating thermal strain and thermally isolating the CCD from the structure. The G-10 mounts have titanium post flexures on either end. Additionally, the OTA itself is mounted to the K13C baffle tube via titanium isolators, which act to mitigate thermal strain and vibrations and to provide conductive isolation.

The baseline mission requires decontamination of the CCD at temperatures >-18°C before door opening, with decontamination heaters. Decontamination can be achieved at lower temperatures, but long time periods may be necessary to desorb water.

An additional thermal requirement is that the CCD must be maintained at a temperature of < -70 °C while acquiring science data. This requirement is met by mounting the CCD to a bracket which is mounted to the OTA via conductively isolating titanium flexures; the CCD bracket is in turn attached to a gold-coated beryllium S200F conduction bar that is bolted to a gold-coated beryllium S200F thermal radiator whose outside surface is painted white with Aeroglaze A276. Since the radiator is mounted to a separate spacecraft panel from the OTA itself, there must be some compliance to allow motion between the two, provided by a highly conductive 1100 series aluminum alloy S-link.

The LORRI in-flight temperatures were predicted via the creation of a finite difference thermal model that included all conductive and radiative heat transfer. A thermal balance test of the instrument was performed that validated the thermal model in five separate test cases. The results of thermal balance testing predict that the 0.5 Watt gradient control heater will not be required in flight. Flight data shows a gradient of 1.5 °C, well within the 2.5 °C requirement for gradient.

### 3.1.4  Telescope mechanical requirements and design

The structural requirements of the OTA were geared towards achieving minimum mass, while maintaining performance over operational temperatures and allowing for the stiffest design that will survive the launch environment. The requirements were a maximum mass of 5.64 kg, a minimum resonant frequency of 60 Hz, and survival under launch-induced vibration and stresses.

The inherently high stiffness-to-weight ratio of SiC, (~4.5 times that of aluminum) allowed the fabrication of a low-mass structure with a light-weighted primary mirror of open back, hub-mounted design to minimize weight. The main baffle tube was fabricated from graphite composite (K13C2U, M55J, and T300), another very high stiffness-to-weight material (~2.5 to 4 times that of aluminum). The smaller internal baffles were fabricated from a light-weight magnesium alloy (ZK60A).

Vibration isolation was required to survive the launch environment. Titanium isolators were incorporated to mount the structure to the main baffle tube at three points approximately at its center of gravity location. Another important structural design consideration was to minimize any potential for mount-induced distortion of the optics. Intimately connected to this design consideration is the requirement for the OTA to mount to a surface with only moderate mounting coplanarity, namely, a spacecraft aluminum honeycomb panel. To avoid degradation of optical quality, a 3-point mount was adopted, with the bases of each of the three mount locations on the OTA outfitted with a ball joint that can be loosened and re-tightened if necessary.

At the interface between the main baffle tube and the OTA inner assembly, the vibration isolators also act to mitigate any mount-induced strains. The flexurized mount plate at the secondary mirror serves the dual purpose of controlling thermally induced distortions, as well as mount-induced distortions; the CCD flexure mounts also serve this dual purpose.

Due to its mass, the primary mirror was not flexure-mounted, as the low resultant frequency and dynamic responses would have increased the risk to the OTA under vibration. Instead the primary mirror is hub-mounted, with a post and a foot bolted via three invar inserts to a mount plate. The mount plate is in turn bolted to the structure. Because the lens cell is made of Invar and mounts to the structure, close to the primary mirror, the mount plate helps to separate any induced thermal strain in the structure from being transferred to the primary mirror.

### 3.1.5 Instrument integration, focus and alignment

When the OTA was assembled at SSG, a convex spherical reflector was centered at the focus using interferometry. Metrology, combined with knowledge of the shim size used to connect the reflector to the carrier plate, determined the location of the focus relative to the interface location on the carrier plate. SSG also provided a reference mirror on the back of the secondary mirror mount, such that the telescope line-of-sight was parallel to the reference mirror's normal. Two optical reference flats at right angles to the line-of-sight were mounted to the LORRI metering structure for use in alignment monitoring after mounting to the spacecraft.

The depth of focus for LORRI at the detector, with mechanical tolerances on the CCD, allowed initial shim sizes to be selected for system focusing at the SSG-provided focus location. A 300 mm aperture, f/5 off-axis parabolic collimator was used to project a point-like image into LORRI. This image was produced by a laser unequal path interferometer (LUPI) at the collimator focus. A series of exposures were made at nine points in the LORRI field by moving within the field using a fold mirror. At each location, the spot was centered on a detector element by viewing the live image from the CCD through ground support electronics, and balancing the wings of the image symmetrically about the center detector element. This was repeated for seven focus adjustments of the collimator, with slight measured changes from nominal to allow for deterministic shimming of LORRI. Data from these exposures were examined to find the best LORRI focus versus the collimator adjustment. New shims were then installed to move the plane of the CCD onto the plane of best focus.

Once at nominal focus, the CCD was centered on the optical axis by use of a theodolite viewing both the reference mirror on the back of the secondary mount and the primary mirror. The theodolite was autocollimated on the reference mirror, and the azimuth and elevation recorded. The theodolite then viewed the four corners of the CCD off the primary mirror, and these azimuth and elevation values were recorded and averaged. The CCD location was shifted in the plane of best focus such that the average of the four CCD corners was within tolerance of the normal to the reference mirror.

After the best focus was found in ambient conditions, LORRI was installed in the NASA Goddard Space Flight Center's Diffraction Grating Evaluation Facility for a focus check at flight-like thermal and vacuum conditions. This was done by viewing a collimated beam, which projected a small point-like image into LORRI, and stepping the spot over the field by tilting LORRI on a gimbal platform.

Prior to delivery to the spacecraft, LORRI's line-of-sight was measured relative to the optical reference flat mounted to the back of the secondary mirror support. Additionally, the roll angle was measured by viewing the CCD corners, and referencing to the orthogonal alignment mirrors. These data, combined with measurements performed referencing the flats to the spacecraft coordinate system, showed that LORRI's line-of-sight was within mission requirements. Tracking of the LORRI alignment references through the spacecraft environmental test program did not show any significant movement relative to the spacecraft coordinate system.

### 3.1.7 Contamination control

The LORRI telescope assembly remained under a nitrogen purge during all phases of integration and test until launch, except for limited times when images were taken or when put under vacuum. The internal cleanliness requirements for the LORRI OTA per the LORRI performance specification are a beginning-of-life specification of 250 A/2 and an end-of-life specification of 300 A per MIL STD 1246C. The resulting predicted reduction in optical efficiency (Conard et al. 2005) is < 4%.

## 3.2    Electronics

As shown in Figure 3, LORRI electronics consist of the ASE and FPU. The ASE contains three printed circuit cards. These are the low voltage power supply (LVPS), the event processor unit (EPU), and the imager input/output (IM I/O). The ASE is the primary interface between the spacecraft and the FPU, which mounts and controls the CCD. Additional information can be found in Conard et al. (2005).

### 3.2.1  Focal plane unit

The LORRI FPU is required to read out a complete image in 1 second, with the charge level in each pixel represented by a 12 bit binary word. A highly sensitive CCD with antiblooming was required, leading to the choice of the E2V Technologies CCD47-20. This is a 1024×1024 pixel frame transfer CCD with 13 micron square pixels. This device is back-illuminated for high quantum efficiency and has a frame transfer time of 13 milliseconds. The FPU noise is required to be <40 electrons per pixel, well above the CCD read noise which is calculated to be about 10 electrons at the readout time of about 0.7 microsecond per pixel. In general, exposures of 50 to 200 milliseconds are typical for LORRI, although the maximum exposure time is 29.9 seconds. The FPU design includes switchable 4×4 pixel on-chip binning. The FPU also includes two small incandescent bulbs that can illuminate the CCD through multiply scattered light, so that testing can be performed even when imaging is not possible through the optics (e.g., when the cover door is closed).

All CCDs have a number of clocks that must be driven to specific levels for satisfactory operation. The E2V CCD uses three phase clocks for image zone, memory zone and line transfer. These clocks are highly capacitive, particularly for the image and memory areas of the chip, and they have capacitive coupling between different phases. LORRI uses Micrel MIC4427 drivers which are designed to drive high capacitance loads at the required voltage levels from logic level inputs. They are switching, not linear, devices so that low and high voltage levels are obtained by suitable choice of supply voltages, and transition rates must be adjusted at the output. This is done with series resistance which adds to the internal switch resistance of the drivers, forming a simple time constant with the capacitance of the CCD phase. The CCD requires 29 volts bias for the output field effect transistor. A charge pump with pre- and post-regulation was used to generate this voltage.

The LORRI FPU uses an Analog Devices AD9807 integrated circuit that performs correlated double sampling, signal amplification, and analog to digital conversion to 12 bits at maximum rates of 6 MHz, comfortably above the pixel readout rates which are close to 1.5 MHz. The AD9807 is susceptible to latch-up from ionizing radiation in space, and the LORRI FPU incorporates latch-up protection circuitry. The CCD output is low enough for the AD9807 amplifier to contribute significant noise, so a low noise, wide band operational amplifier is added between the CCD and the correlated double sampler, and the AD9807 is run at low gain.

### 3.2.2 Associated support electronics

The LORRI EPU controls the instrument via interfaces to the LVPS and IM I/O boards. The EPU communicates to the spacecraft using an RS-422 link, which receives commands and transmits engineering data. The EPU uses a RTX2010RH processor and runs FORTH code.

The main function of the IM I/O board is to receive serial image data from the FPU and transmit that data to either of two Integrated Electronics Modules (IEMs) in the required format. There are dual, redundant IEMs on the *New Horizons* spacecraft, which provide command, data handling, and telemetry functions. One IEM is active and one is a back-up; LORRI provides both interfaces (LVDS and RS-422) to both IEMs. Secondary functions of the LORRI IM I/O board include the ability to:  store and transmit the image header, receive commands from the RTX processor, calculate a 32-bin histogram, generate test patterns without an FPU present, command the FPU mode and exposure times based on input from the RTX.

The Imager I/O board contains two field programmable gate array (FPGA) designs. The first is called the imager-interface FPGA, and the second is called the RTX-bus FPGA. The main function of the imager-interface FPGA is to read images from the FPU and send them to the IEM. As an added feature, the IM I/O board can generate test pattern images across the IEM interface without an FPU present. The first pattern consists of a horizontal ramp and the second pattern consists of a vertical ramp. The imager interface FPGA can also receive data from the RTX. These data are sent across the ASE backplane and through the RTX-bus FPGA. The data are used to set the FPU mode and exposure time, to set the active IEM low voltage differential signaling port and to write the 408 bit header. The FPU mode data is transmitted to the FPU across the pixel data signal at the beginning of each second. The header data replaces the first 34 pixels when data is sent to the IEM, to provide redundancy in associating instrument engineering data with each image. The instrument engineering data are transmitted by the spacecraft within a separate data stream from the science data, and the engineering data must be associated with individual images in ground processing. For LORRI, the critical header information is also encoded into the images themselves, at the cost of 34 pixels in the first row of each image.

The RTX-bus FPGA also calculates a 32-bin histogram of the FPU image data currently being transmitted. This histogram is then made available to the RTX for future exposure time calculations. The RTX FPGA also collects the FPU status and temperature data, making it available to the RTX.

The LVPS provides 2.5 V, 6 V, and 15 V power as required by the other boards within the ASE and by the FPU. The input voltage from the spacecraft is 30±1 V. The LVPS also provides for current, voltage, and temperature monitoring via an $I^2C$ serial interface to the EPU board. The LVPS board provides switching to control power on/off to the FPU and the telescope trim heaters.

### 3.2.3 Flight software

The LORRI RTX-2010 processor shares a common design with that of the PEPSSI instrument on *New Horizons* (see accompanying paper). This common design extends into the software. The common flight software provides packet telemetry and command handling services. Besides handling LORRI-specific packets, the common software automatically generates a variety of standard packets, including housekeeping/status, command echo, memory dump, and alarm packets. Similarly, besides handling LORRI-specific commands, the common software also handles standard commands for memory loads and memory dump requests. The command handling software also provides storage and execution of command sequences. The common software has timekeeping, voltage and current monitoring, and memory management services and a standard boot program.

The LORRI-specific flight software controls heaters, collects voltages, currents, and temperatures from the LVPS board, and manages the FPU. In the FPU, the software controls the exposure time, either by manual command or automatically based on the hardware-provided image histogram, generates an image header, and enables routing of the image to the spacecraft. The software also controls the FPU's test patterns and calibration lamps. Whenever the software has nothing to do, it reduces the processor's clock rate to save power.

## 3.3    Laboratory Test and Calibration

The LORRI instrument was subjected to an environmental qualification program. Performance and environmental testing was performed at both the subassembly and instrument levels. Typically, performance tests or calibrations were done before and after environmental testing.

### 3.3.1    Subassembly Test and Calibration

The OTA was tested by SSG. Wavefront testing was performed using a LUPI operated in double-pass mode. Wavefronts were measured before and after vibration test and during thermal vacuum test. No change was detected due to the vibration test. Some change was noted at cold temperature during thermal vacuum testing, but it was determined that the level of change was acceptable within the LORRI performance requirements. As a secondary verification of performance, modulation transfer function testing was also performed by projecting small features into the OTA, and recording highly magnified images on a non-flight detector. This testing was performed only in ambient conditions and showed excellent correlation to wavefront data.

The FPU and CCD were also subjected at JHU/APL to testing at the subassembly level. They were calibrated at predicted operating temperatures and then subjected to environmental testing. The calibration of the FPU and CCD was not repeated after the environmental series due to time limitations.

### 3.3.2    Instrument Level Calibration

Results of the LORRI instrument-level laboratory calibrations are discussed in detail by Morgan et al. (2005). The instrument-level optical calibrations under ambient conditions were performed at JHU/APL. Due to schedule conflicts with another instrument development program at JHU/APL, LORRI instrument-level optical calibrations in thermal vacuum were performed at the NASA Goddard Space Flight Center, in the Diffraction Grating Evaluation Facility (DGEF).

The objective of radiometric calibration is to determine the conversion from the observed signal in instrumental units to the scene radiance in physical units. For LORRI, the radiometric calibration equation relates the digital output $S$ expressed as DN (12-bit 'data numbers') to the scene radiance (or surface brightness) $I$

$$I_{x.y.\lambda} = \frac{S_{x.y.\lambda.T.t} - Bias_{x.y.T} - Dark_{x.y.T<t} - Smear_{x.y.\lambda.T.t.\dot{o}} - Stray_{x.y.\lambda.T.t.\Phi}}{FF_{x.y.\lambda.T} \ R_{\lambda.T} \ t} \ . \tag{1}$$

where subscripts $x; y$ denote dependence on detector position (i.e., line and pixel numbers), $\lambda$ on wavelength, $T$ on temperature, $t$ on exposure time, $\phi$ on optical power at other points within the FOV, and $\Phi$ on total optical power within the telescope. The quantities and their units are: $I$ the scene radiance in W m$^{-2}$ sr$^{-1}$ nm$^{-1}$; $S$ the observed signal in DN; *Bias* the electronic offset of the CCD signal in DN; *Dark* the detector dark current in DN; *Smear* the signal in DN acquired by a pixel as it is shifted through the scene during frame transfer; *Stray* the signal in DN due to stray light; *FF* the (dimensionless) flat field response which normalizes responses of individual pixels using a uniform diffuse source; $R$ the absolute responsivity in [DN s$^{-1}$pixel$^{-1}$]/[Wm$^{-2}$sr$^{-1}$nm$^{-1}$]; and $t$ the exposure time in seconds.

Of the terms in the radiometric calibration equation, bias and smear were characterized over the relevant parameters; dark is insignificant at operational temperatures; stray light was tested at discrete off-axis source angles; flat field response was measured, but there were difficulties obtaining the desired 0.5% accuracy at all pixels from ground-based data, and in-flight measurements are planned to replace or correct the current flat field; and absolute response was measured with sufficient accuracy for LORRI science goals. In addition to radiometric calibration, instrument characteristics such as detector read noise, the imaging point spread function (PSF), effective focal length and field-of-view were determined during calibration. The PSF was measured at several locations across the FOV, and the field-of-view and read noise were determined.

### 3.3.3.    Calibration setup

To confirm stability of LORRI's focus and response, key measurements were conducted at the nominal operating temperature as well as temperatures at the upper and lower ends of the operational range. The telescope assembly temperature ranged from -97°C to -60°C during calibration, while the CCD temperature ranged from -96°C to -78°C. At these low temperatures it was necessary to operate LORRI in vacuum to avoid condensation. The calibration was conducted at the NASA/GSFC DGEF which operates as a large vacuum chamber for optical calibrations. Calibrations at DGEF were conducted in July, 2004 prior to instrument-level environmental (vibration and thermal vacuum) testing of LORRI, and again in September, 2004 after environmental testing, to establish stability of the calibration.

Figure 5 is a block diagram of the DGEF calibration setup for LORRI. A light source outside the vacuum chamber was fiber optically coupled into the chamber, feeding an integrating sphere. The integrating sphere backlit a rotating target wheel that contained various apertures such as pinholes and resolution targets. The target wheel was at the focal plane of a 38 cm diameter collimator, the output of which illuminated LORRI, which was mounted on a two-axis gimbal approximately two meters from the collimator aperture. Between LORRI and the collimator, a 30.5 cm diameter off-axis paraboloid (OAP) reflector could be rotated into the beam to focus a pinhole image onto a calibrated photodiode, providing a measurement of the integrating sphere port radiance for radiometric calibration.

A 150 Watt, high pressure, ozone-free xenon arc lamp was used for the LORRI calibration in order to provide a high radiance, solar-like spectrum. Since there are significant spectral variations in response over the LORRI bandpass, it is important to approximate the expected spectral radiance distribution in order to infer broadband radiometric response from calibration measurements with reasonable accuracy. Pluto, Charon, Jupiter, and Jovian satellites have spectra ranging from generally neutral to moderately red relative to a solar spectrum.

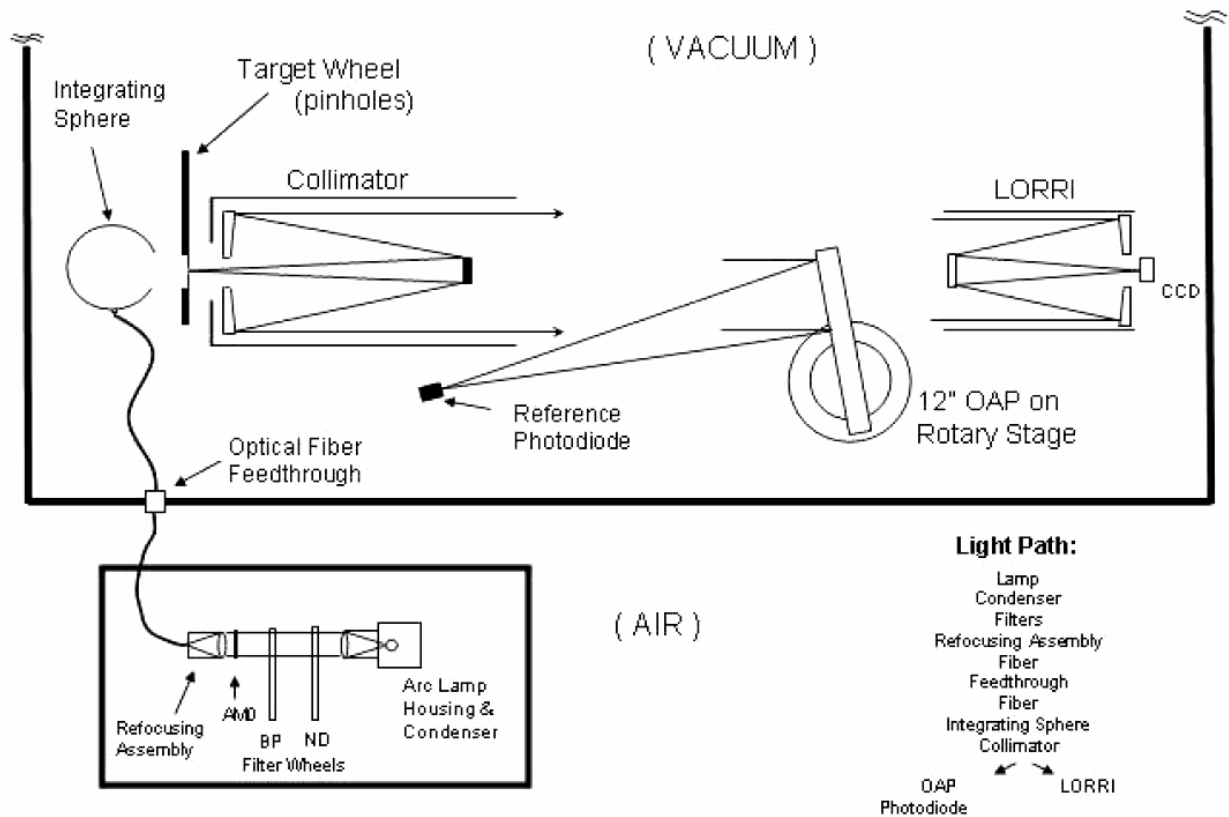

**Figure 5. Calibration Facility at the NASA Goddard Space Flight Center**

The arc lamp output was collimated using a fused silica condenser and transmitted through two six-position filter wheels, one with bandpass (BP) filters and the other with neutral density (ND) filters. The output then passed through an airmass zero (AM0) filter and was refocused into a fiber optic providing input to an integrating sphere inside the vacuum chamber. The AM0 filter attenuates the near-infrared portion of the spectrum to make the overall spectral energy distribution more solar-like. The BP filters were centered at 400 nm, 500 nm, 600 nm, 700 nm, and 850 nm. Bandpasses at full width half maximum (FWHM) for the filters were 65 nm for the 400 nm filter, 119 nm for the 850 nm filter, and approximately 90 nm for the others.

The radiance at the integrating sphere output port, spectrally integrated over the LORRI bandpass, was 46 $\mu$W cm$^{-2}$ sr$^{-1}$. Figure 6 shows the source spectrum. Although the lines appear to dominate the spectrum in Figure 6, they contribute a modest fraction of the total energy in the spectrum because their widths are relatively narrow. The calibration source spectrum is compared to the spectra of Pluto and Jupiter, as well as LORRI's calculated spectral response in Figure 6 (Morgan et al. 2005).

The target wheel (Figure 5) contained image masks, including pinholes ranging in size from 5 $\mu$m to 1000 $\mu$m, an NBS 1963A resolution target, and an open position. The target wheel was located at the focal plane of a Cassegrain collimator with a 38 cm aperture, 460 cm focal length, and 0.5° unvignetted field of view. The collimator aperture was large enough to overfill the LORRI aperture without vignetting over the LORRI FOV and gimbal motion range, with tolerance for easy co-alignment. The two-axis gimbal was stepper motor-driven with a Unidex controller at 63 steps per arcsecond to scan LORRI across targets. A LORRI pixel subtends roughly one arcsecond. The reference photodiode was a Hamamatsu S1336-8BQ, whose responsivity was calibrated by the vendor at 10 nm intervals from 200 to 400 nm, and at 20 nm intervals from 400 to 1180 nm.

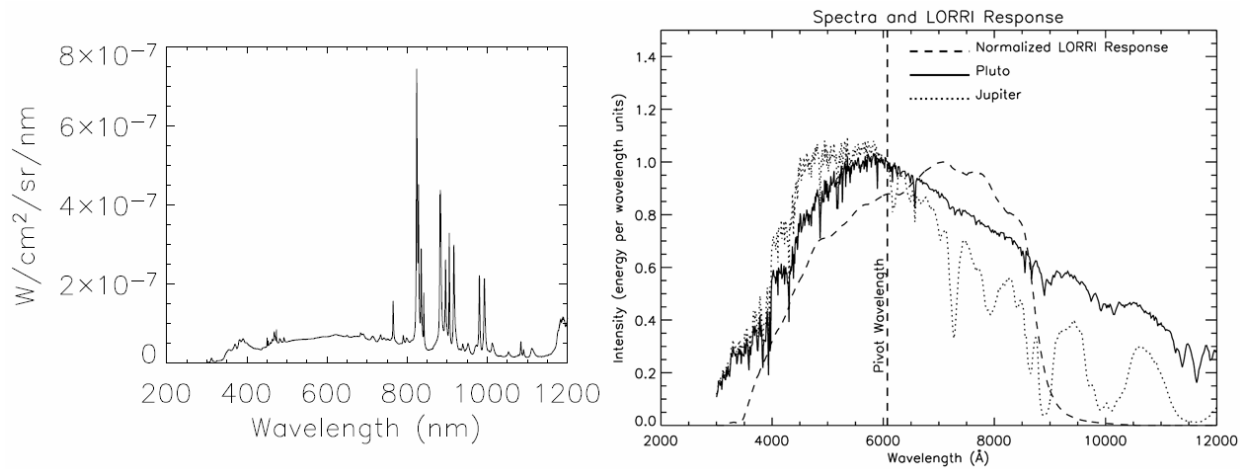

**Figure 6. (left) Arc lamp calibration source spectrum and (right) Spectra of Pluto, Jupiter normalized to unity at 607.6 nm and LORRI response normalized to unity at maximum (Morgan et al. 2005)**

For thermal control during calibration, a shroud cooled by liquid nitrogen was constructed surrounding the telescope. A small annular shroud encircling the telescope inside the cold shroud was heated to control the telescope temperature to the desired set point. The thermal conductivity of the telescope structure was sufficient to guarantee a uniform temperature distribution despite the uneven heating. The shrouds were fixed, with LORRI gimbaled inside them. LORRI viewed the collimator through a ten inch aperture in the main cold shroud, which kept the shroud edge well out of the LORRI FOV over the maximum gimbal motion range used. LORRI calibration temperatures are listed in Table 7.

**Table 7. LORRI calibration temperatures**

|  | Nominal | Cold | Hot |
|---|---|---|---|
| CCD | -80°C | -93°C | -75°C |
| FPU Board | 26°C | 10°C | 35°C |
| Primary mirror | -72.5°C | -98.9°C | -62.1°C |
| Secondary mirror | -73.3°C | -99.3°C | -63.5°C |

### 3.3.4. Laboratory calibration results

Results of pre-flight calibrations are summarized here, and preliminary analyses of in-flight calibrations from instrument commissioning are summarized in the next section. Pre-flight and in-flight calibrations are fully consistent, considering limitations of ground test equipment.

The detector read noise was estimated from zero exposure time exposures taken in the DGEF at nominal temperature with no illumination. One hundred of these dark images were acquired, and the standard deviation of the signal of each pixel was evaluated. The distribution of the standard deviations is plotted in Figure 7. The most probable standard deviation is 1.2 DN, which is adopted as a read noise estimate.

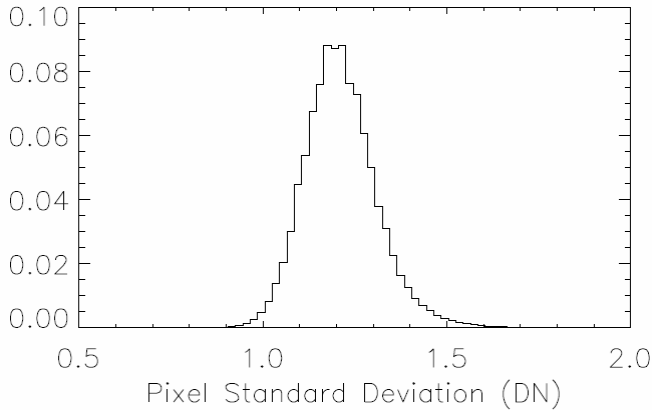

**Figure 7. Read noise distribution, from 100 dark images measured pre-flight. The estimated read noise is 1.2 DN.**

The detector bias (signal at zero illumination due to electronic offset) depends on the focal plane electronics board temperature. From subsystem level tests, the bias in unbinned 1×1 mode is

$$Bias \text{ [DN]} = 509.941 \ (\pm 0.151) + 1.073 \ (\pm 0.005) \times T \tag{2}$$

in units of DN as a function of board temperature $T$ in °C. The observed in-flight board temperature range is +34 (±0.7)°C. The bias ranges from 546 to 547 DN over that range. The bias in 4×4 binned mode is 2 DN higher. Detector dark current is negligible over the flight CCD temperature range, for exposure times up to at least one second. Planned exposure times are on the order of 100 ms. Also from subsystem level testing, the FPU gain is 22 (+0,-0.5) electron/DN for CCD temperatures of -50°C and -70°C and for board temperatures of 0°C and 40°C. This was determined from analyses of photon counting noise in flat field exposures using constant illumination with a green light emitting diode. A gain estimate of 22±0.4 electron/DN was also obtained with analyses of noise in flat field exposures versus net signal with eight different exposure times. The linearity of the FPU was found to be within ±1% over DN values from 900 to 3900, where the saturation level is 4095.

During calibration at the DGEF, point source imaging performance was evaluated by observing a 3×3 grid of 5 μm pinholes. The grid target was arranged with 4 pinholes near the corners of the FOV, 4 along the edges, and one near FOV center. LORRI was gimbaled through a 6×6 subpixel grid of positions spanning approximately 1.5 pixels, acquiring 10 images at each step. A fitting process was applied to the images to estimate the sub-pixel irradiance distribution. Results indicated a PSF with FWHM on the order of 2 pixels, with no discernible dependence on position within the FOV. No significant variation of the FWHM with temperature was observed, verifying focus stability from ambient temperature to -100°C. These tests were adversely affected by diffraction by the collimator secondary obscuration, which was larger than the LORRI secondary.

More accurate PSF measurements were obtained in bench tests at ambient temperature at APL. A source was projected through a Laser Unequal Pathlength Interferometer (LUPI) and then an off-axis parabola collimator with an unobstructed aperture to generate a collimated beam that completely filled LORRI's aperture. For LORRI tests, the beam was reflected from an optical flat mounted with tip/tilt controls, which allowed focus and PSF measurements at five positions in the LORRI FOV (near the corners and near the center). Image analysis indicated a PSF with FWHM of 1.5 pixels with little variation across the FOV, when fit with a two dimensional Gaussian function. With the spot near pixel center, ensquared energy in a pixel exceeded 0.3 at all four corners and the center of the CCD.

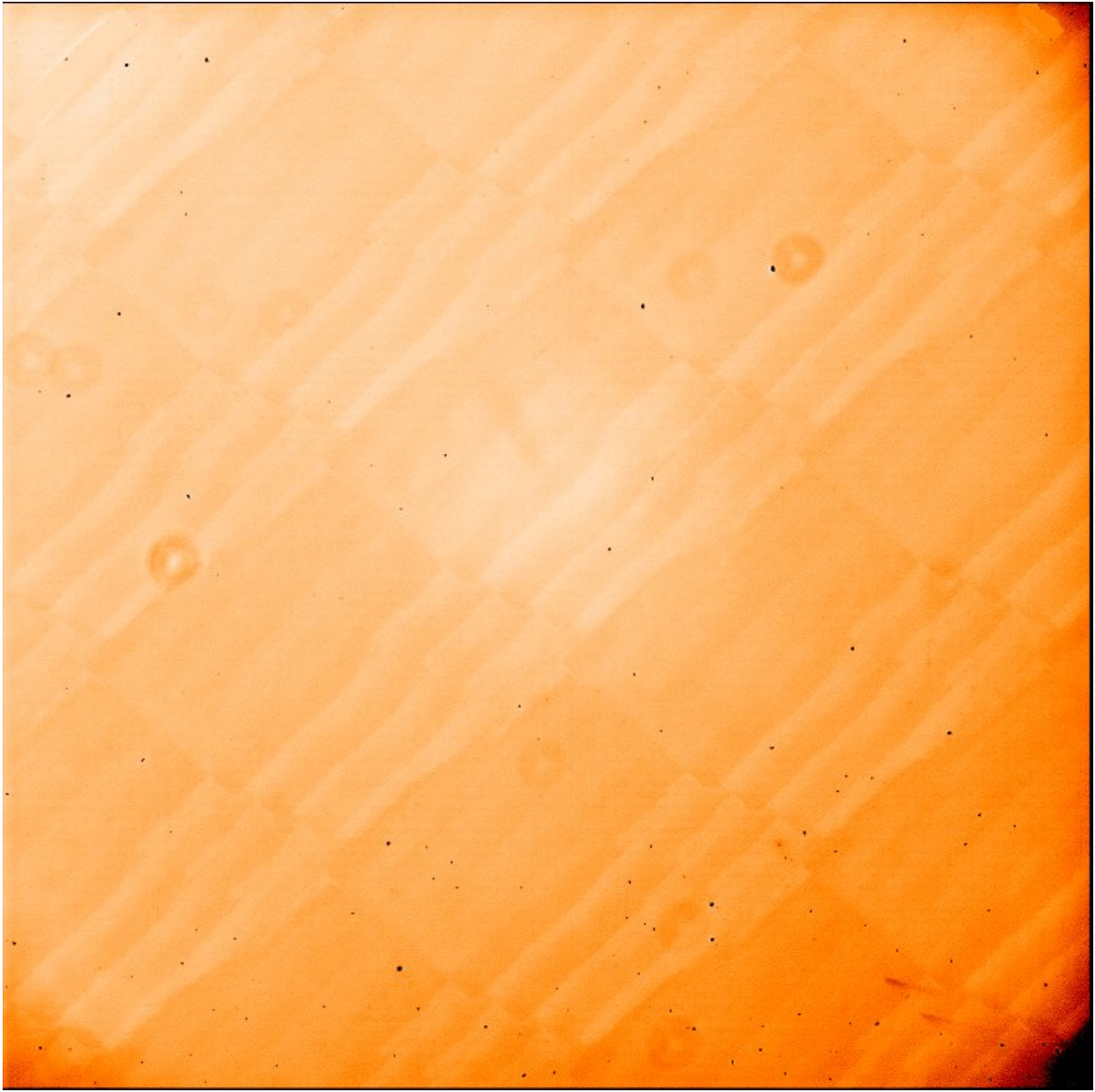

**Figure 8. Provisional flat-field measured at APL: logarithmic stretch from 0.95 (black) to 1.05 (white). Dust particles form 'donuts' and dark spots. There is 4% vignetting in the corners, but in the upper left corner and in the center of the field, there is excess brightness from image ghosting which is not removed in preliminary processing.**

Although LORRI flat field images were obtained during the post-environmental calibration at the DGEF, the CCD was subsequently cleaned by blowing ionized nitrogen across it. New flat field measurements conducted at APL with a quartz halogen lamp (see Figure 8) showed that the number of pixels affected by particle shadowing was reduced to 62 from over 150. These new flat field measurements were conducted in air at room temperature. There is vignetting of approximately 4% in the corners, consistent with expectations based on the design of the baffles and field stop. Low amplitude, dark "doughnut" rings (up to 1% amplitude) are seen from particles probably on the field-flattening lenses, and dark spots (typical 20% signal loss in single pixels) are seen from particles on the CCD. Comparison with in-flight images of Jupiter and with calibration lamp images shows that the particles have not moved since pre-launch calibration, and no new particles have been introduced since launch.

A central region of approximately 200 pixel radius is brightened by a maximum of 1.5% by ghost images that result from multiple reflections between surfaces in the field flattening lens group. There is also excess diffuse brightness from image ghosting in the upper left corner of Figure 8 such that the known vignetting is masked, and arc-like ghost features have been removed for preliminary processing. The ghosts are dominated by out-of-field illumination at the red extreme of the LORRI passband, depending on the radiance distribution over field angles just outside the FOV up to approximately 0.37° off-axis. The ghosts are strongly dependent on source spectrum and will be characterized extensively with Jupiter observations. Also seen in Figure 8 is fringing of 0.5% amplitude from constructive and destructive interference in the CCD.

The DGEF flat fields showed no significant temperature dependence. The maximum differences between the flat field at nominal temperature and those at the hot and cold extremes were on the order of 1%, with fewer than 250 pixels exceeding 0.5% variation in either case. Bandpass-filtered flat fields showed somewhat more variation with temperature, but for most of the filters the variations are not statistically significant. Only 10 images were used for the filtered flats, compared to 100 for the panchromatic flats. There appeared to be significant temperature dependence in the 850 nm flat, possibly due to interference in the CCD with the emission lines in the xenon arc lamp spectrum. These strong lines do not occur in natural targets, and the panchromatic flat fields are practically independent of temperature for flight conditions.

However, laboratory flat field calibrations apply only imperfectly to the flight system, because of limitations such as use of non-solar source spectra. The provisional flat field of Figure 8 will be replaced with in-flight flat field observations, using Jupiter as an extended source during the Jupiter flyby, either to use directly or to be combined with the laboratory flat field data. Additional in-flight flat field data are planned using solar stray light.

The absolute radiometric response of LORRI was also determined at the DGEF. When system responsivity varies significantly over the spectral bandpass, as is the case for LORRI, the measured signal depends on the scene spectrum, which will depend on the target. As can be seen in Table 3, LORRI has requirements for relative, not absolute, radiometry. However, the source spectra that LORRI will view in flight will be measured by other instruments on New Horizons (namely Ralph, see companion papers). With the aid of independent spectral observations, LORRI can provide useful radiometric data.

LORRI's absolute responsivity as a function of wavelength was determined (Morgan et al. 2005) using panchromatic absolute measurements together with a calculated relative response curve. First, the LORRI responsivity spectrum was calculated from geometrical throughput of the optical design and from spectral characteristics of optical components and the CCD. It was assumed that an unknown constant factor would scale this response spectrum to the correct absolute level. The calculated spectral responsivity accounted for the measured mirror reflectance curves and lens transmissions, and it used E2V's typical quantum efficiency curve for the model 47-20 CCD which was not individually measured for the flight CCD.

Second, the absolute LORRI response was estimated by scaling the calculated spectral responsivity to match the DGEF flat field observations. The flat field source spectrum was corrected to an absolute scale using reference photodiode measurements taken immediately before each set of flat field images. The 1 mm pinhole was imaged onto the calibrated photodiode with the OAP mirror to obtain the spectral radiance of the integrating sphere port (see Figure 6). Correction was made for signal reduction from the partial blockage of LORRI's aperture by the collimator secondary, which is larger than LORRI's secondary. The LORRI obscuration by spider and secondary is 11%. The LORRI signal was taken from the median flat field signal, which was 15908 DN/s/pixel viewing this diffuse source spectrum. From this signal, the absolute LORRI response curve shown in Figure 9 was derived (correcting the figure shown in Morgan et al. 2005).

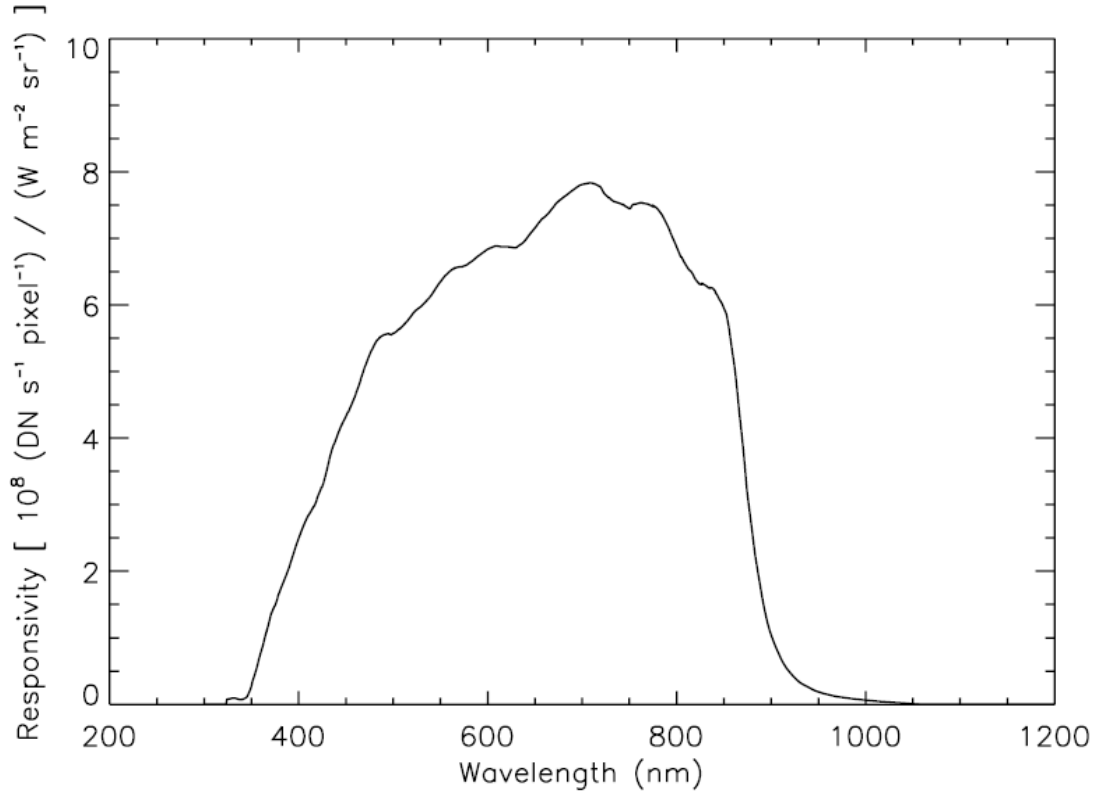

**Figure 9. Absolute monochromatic response from calculated spectral responsivity, correcting Morgan et al. (2005).**

Lastly, the response to realistic scene spectra was calculated numerically by integrating the product of the assumed scene spectra and the LORRI response spectrum derived from calibration. From the calibrated response curve, LORRI's responsivity to specified scene spectra can be calculated by integrating the product of the LORRI absolute response curve and the normalized scene spectrum. For a source with the average Pluto spectrum, the absolute response of LORRI is estimated from the DGEF calibrations as

$$R_{\lambda, T}\left(607.6\,\text{nm}\right) = 2.2 \times 10^{11} \frac{\text{DN/s/pixel}}{\text{W/cm}^2/\text{sr/nm}} \qquad (3)$$

giving the conversion from spectral radiance at the LORRI pivot wavelength (Horne, 2004), which is 607.6 nm, to signal expressed in units of DN/pixel/s. The spectral radiance in the denominator is understood to correspond to the pivot wavelength. The absolute response depends on the shape of the adopted scene spectrum; for additional details see Morgan et al. (2005) and discussion of in-flight observations below.

## 3.3.5 In-Flight Calibration

LORRI has successfully completed its in-flight commissioning tests. LORRI's cover door opened on August 29, 2006, and successful observations have been acquired of the open star cluster Messier 7 and the prime mission target Pluto as well as the planets Uranus, Neptune, and Jupiter. Instrument commissioning tests through September, 2006 are summarized in Table 8.

**Table 8. LORRI In-Flight Commissioning and Calibration Test Summary**

| Dates and description | Sequence name |
|---|---|
| 24 Feb 2006: all currents, voltages, power, and temperatures nominal | LORRI-005 |
| 23-24 April and 2-3 May 2006: in-flight noise test | LORRI-025 |
| 30 Jul 2006: Dark series bias images, cover door closed | LORRI-006 |
| 29 Aug 2006:  Door opening verification images (Messier 7) | LORRI-007 |
| 31 Aug 2006:  Radiometric/PSF images (Messier 7) | LORRI-010 |
| 31 Aug 2006:  MVIC co-alignment images | LORRI-018 |
| 10,19 Sep 2006: Uranus | LORRI-029 |
| 10,19 Sep 2006: Neptune | LORRI-030 |
| 04 Sep 2006: Jupiter exposure/auto exposure/scattered light tests | LORRI-027 |
| 21, 24 Sep 2006: Pluto | LORRI-023a |

The commissioning test sequences LORRI-005, LORRI-006 and LORRI-025, with the door still closed, were used to verify the FPU noise and bias characteristics in-flight versus the ground calibrations. Results of the in-flight tests have been consistent with the ground calibrations of the FPU. For instance, the difference of two consecutive zero-exposure time images, obtained from LORRI-005 on February 24, 2006, yielded a read noise estimate of $1.11 \pm 0.12$ DN, where DN $\sim 22$ e$^-$. The cosmic ray "hit rate" was also measured from LORRI-005, with the rate of hits (mean ± standard deviation) determined as $15.7 \pm 13.3$ per frame for nominal exposures <0.5 s, where a "hit" was defined to be a pixel with DN at least 5 standard deviations above the mean.

The result of preliminary in-flight measured bias correction, from analysis of 25 frames acquired during the LORRI-006 test performed on 30 July, 2006, is shown in Figure 10. A total of 100 frames were acquired in that test and are under analysis. The preliminary analysis used 25 of the available dark frames, all of which are exposures of zero seconds. For each of these dark frames, the median DN is found of all the pixels in the four dark columns (which are the masked columns at the right side of the image), and the resulting scalar is subtracted from each pixel value in the image zone to obtain a corrected dark frame. After removal of outlier pixel values (more than 3 standard deviations from the mean of the 25 values at the same row and column numbers), the 25 corrected dark frames are then averaged to obtain the so-called *delta_bias* frame which is shown in Figure 10. This frame shows the pixel-level variations of the bias and demonstrates the excellent uniformity of the LORRI flight CCD. The low-level vertical banding in Figure 10 is the result of amplifier oscillation in the FPU; the maximum peak-to-trough amplitude of this wave-like feature is ~0.8 DN.

The *delta_bias* frame is used in the LORRI pipeline data processing to remove the FPU bias from a LORRI image as follows. First, the median of the pixels in the four dark columns of the raw image is determined and subtracted from each pixel in the image zone of the raw image. Then, the *delta_bias* frame is subtracted to obtain the final bias-corrected result. The statistics of *delta_bias* frame are: the mean value in DN is -0.0049; the standard deviation is 0.26; the maximum value is 1.64; the minimum value is -1.56. Since the pipeline processing uses the absolute bias level estimated from the dark columns of every image, the bias estimate from Equation (2) is not used, although the latter provides an independent check.

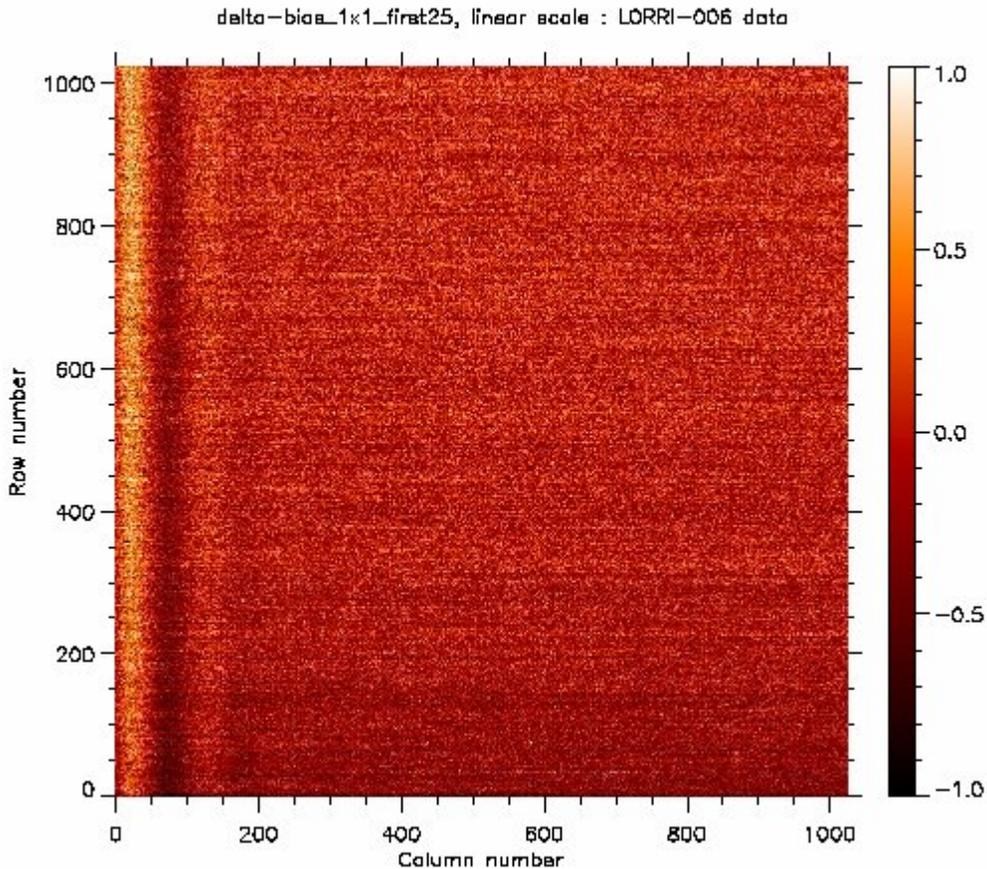

**Figure 10.** *Delta_bias* **frame: in-flight measurement of FPU bias correction, based on analysis of 25 frames acquired 30 July 2006. This image is linearly stretched between -1 DN and +1 DN.**

The next step in LORRI pipeline processing is to apply the flat field correction. Acquisition of in-flight flat field observations is planned during and after Jupiter encounter in 2007. The in-flight observations to date have been reduced with flat fields determined in pre-flight laboratory testing, see Morgan et al. (2005).

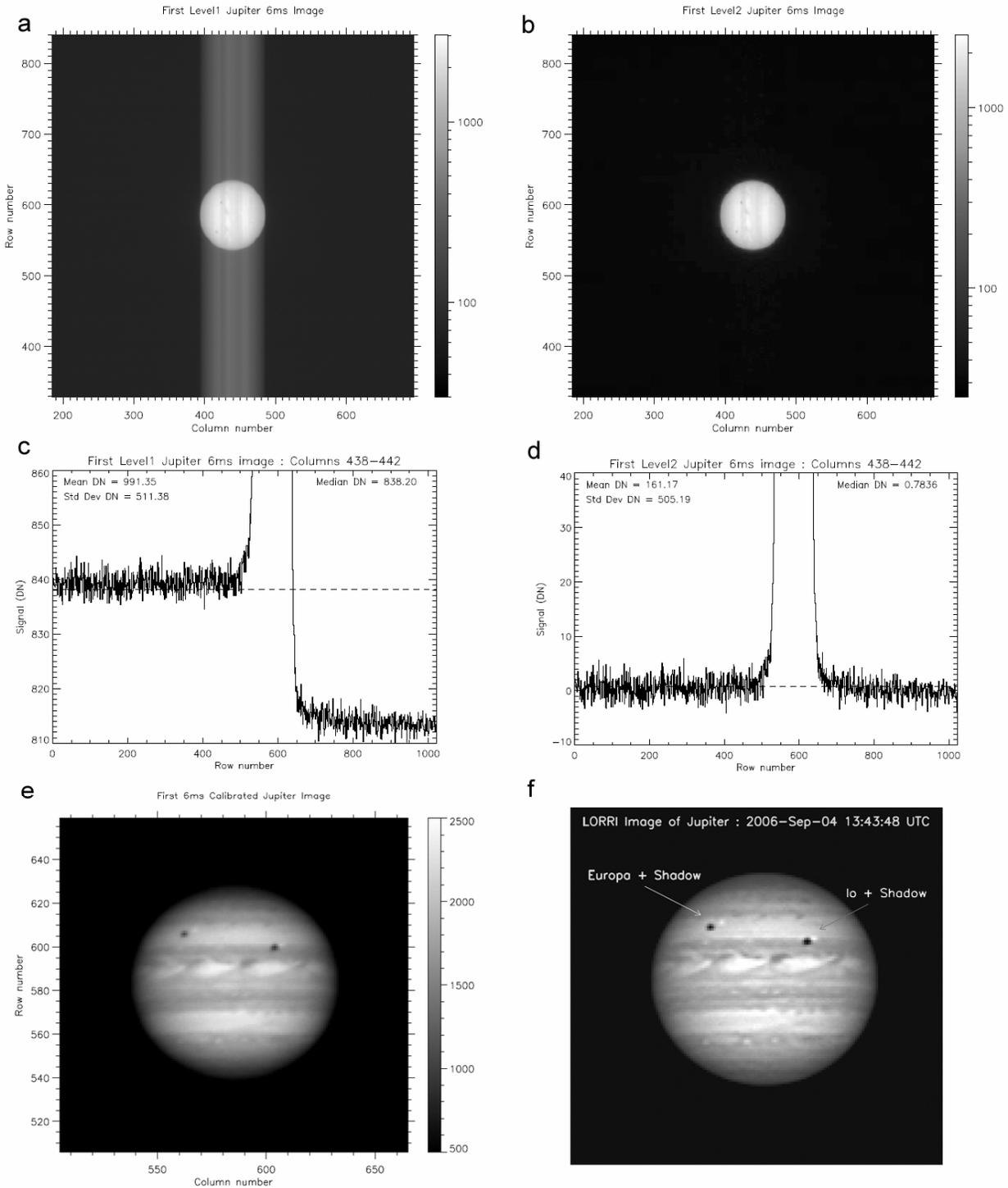

Figure 11. Jupiter observed 9/4/2006. Panel a. Raw image MET 19683344 on logarithmic scale, showing readout smear. Panel b. Same image after calibration on same logarithmic scale, showing bias and smear removal. Panel c. Average of indicated columns showing different readout smear on either side of Jupiter. Panel d. Same as c. for calibrated image, showing removal of readout smear. Panel e. Image of panel a. after

**calibration, on linear scale. Panel f. Five LORRI images including that of panel a., all with 6 ms exposures, after calibration and co-addition. Io and Europa (both barely resolved) and their shadows are indicated.**

Also required to complete the data reduction is removal of an effect called readout smear. This arises from the operation of the frame transfer CCD used in LORRI, where first the image zone is flushed, then an exposure is taken, and finally the image is transferred into the storage zone. Hence a pixel of the raw image is exposed to the scene radiance from the corresponding geometrical element of the scene, but it is also exposed to the radiances of all the scene elements in the same image column during the image transfers. Thus the raw image is the superposition of the scene radiance and the signal acquired during frame transfers, which is called readout smear.

The readout smear is removed as follows. Let $P_{i,j}^{meas}$ = measured image array in DN where $i, j$ are the column and row indices, respectively. Let the exposure time be written $T_{exp}$, with the transfer times for the frame scrub $T_{f1}$ and the frame storage $T_{f2}$ and with $N$ the number of rows (which is 1024 for 1×1 images). Let $T_{favg}$ be the average of $T_{f1}$ and $T_{f2}$ to define the constant

$$A = \frac{T_{\exp}}{T_{\exp} - \dfrac{T_{favg}}{N}} .$$ Finally we define the $N \times N$ constant matrix $\quad \varepsilon_{k,j} = \begin{cases} T_{f1}/T_{favg} & \text{for k < j} \\ 1 & \text{for k = j} \\ T_{f2}/T_{favg} & \text{for k > j} \end{cases}$

with $k, j = 1, \ldots, N$, and we calculate the $N \times N$ matrix $\quad \lambda_{i,j}^{(1)} * T_{\exp} = A\left[ P_{i,j}^{meas} - \dfrac{A\,T_{favg}\sum_k P_{i,k}^{meas}\varepsilon_{k,j}}{N\left(T_{\exp} + A\,T_{favg}\right)} \right].$

The desmeared image is then

$$P_{i,j}^{desmear} = A\left[ P_{i,j}^{meas} - \frac{A\,T_{favg}\left[\sum_k P_{i,k}^{meas}\varepsilon_{k,j} + \dfrac{E_{i,j}}{A}\right]}{N\left(T_{\exp} + A\,T_{favg}\right)} \right]$$

with $\quad \dfrac{E_{i,j}}{A} = T_{favg}\left[ \sum_k \lambda_{i,k}^{(1)}\varepsilon_{k,j} - \dfrac{1}{N}\sum_l \sum_k \lambda_{i,k}^{(1)}\varepsilon_{k,j}\varepsilon_{l,j} \right].$ In-flight tests have verified desmear by this technique using observations of Jupiter obtained at exposure times as short as 1 ms. For the image in Figure 11, $T_{favg}$ = 10.5 ms and $T_{f1}/T_{favg}$ = 1.044, while $T_{f2}/T_{favg}$ = 0.956.

In-flight photometric calibration was obtained using observations of the open star cluster Messier 7 obtained from LORRI-007 on 29 Aug 2006. The visual magnitude is given by

V = -2.5 log $S$ + 18.94 + $CC$

with $S$ in [DN/second]. We find the color correction $CC$ = -0.06 for OB stars, $CC$ = 0 for FG stars, and $CC$ = 0.4 for K stars.

The Messier 7 observations (see Figure 12) also confirm from stellar images that the point source function of the LORRI system, including the effects of spacecraft pointing jitter, is 1.8 pixels FWHM. Stars to at least 12[th] magnitude are detected. The plate scale, based on comparisons of pixel separations of bright stars in the field versus cataloged positions, is 4.96 μrad/pixel. Any geometric distortion in this image is less than 0.5 pixel.

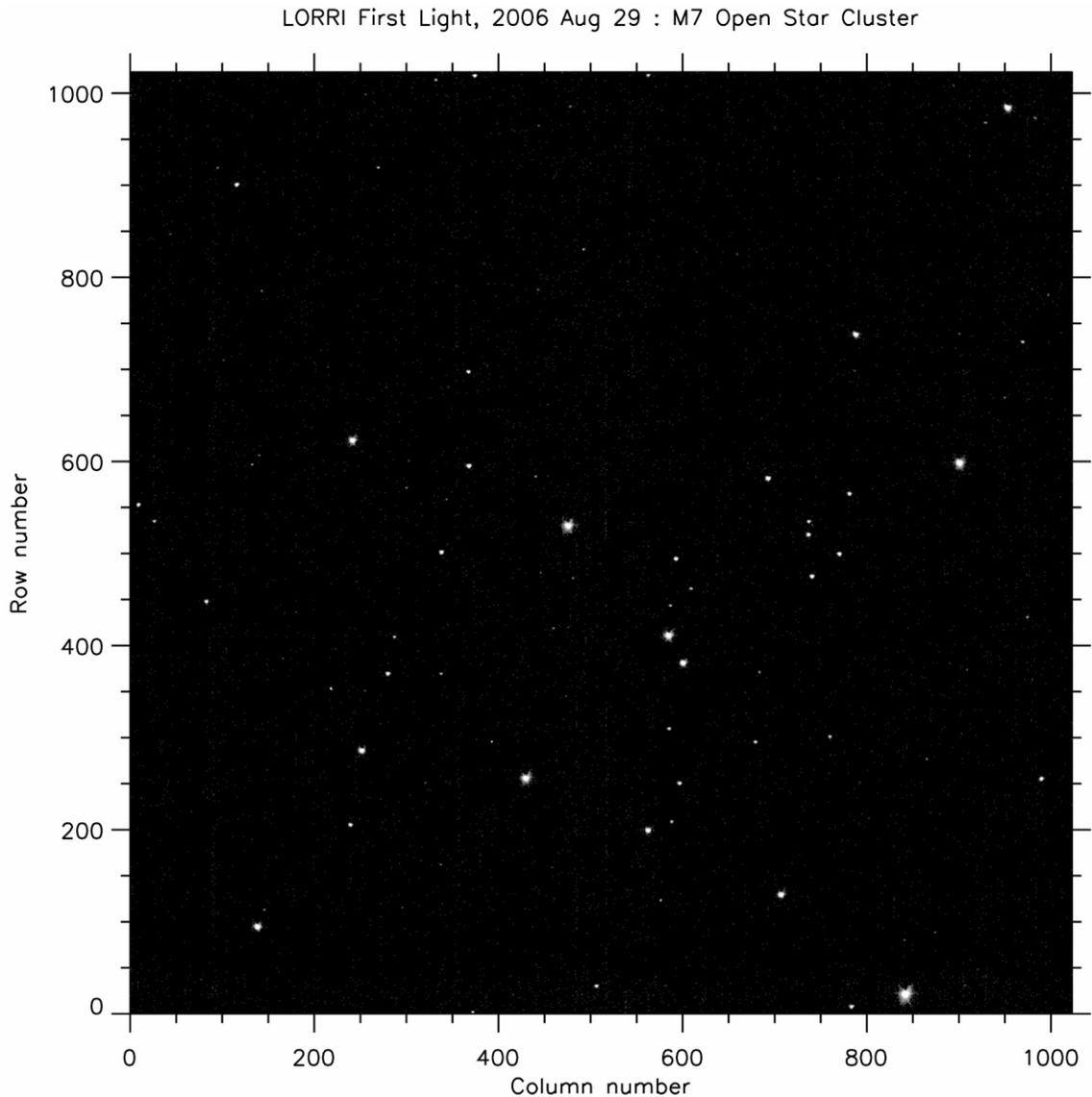

**Figure 12. LORRI image of open cluster Messier 7 obtained in flight. This image has been logarithmically stretched. North is up, east is to the left.**

# 4. Summary

The LORRI instrument was successfully developed, qualified, and calibrated, and it was delivered to the New Horizons spacecraft on time and within budget. It has operated successfully in flight, and its performance is completely nominal. It will obtain the first high resolution imaging observations in the Pluto system and at one or more additional Kuiper Belt objects.

# Acknowledgments

We thank NASA and the New Horizons Project teams for support. We also thank many more members of the LORRI team whose contributions have helped to make LORRI a success.